\documentclass[useAMS,usenatbib]{mn2e}

\usepackage{psfig, epsf, epsfig}

\title[Origin of massive GCs]{Formation of massive globular clusters
with heavy element abundance spread in the Galactic building blocks}
\author[K. Bekki]
{Kenji Bekki${}^1$\thanks{E-mail:
bekki@cyllene.uwa.edu.au} \\
${}^1$ICRAR M468
The University of Western Australia
35 Stirling Hwy, Crawley
Western Australia, 6009}

\begin{document}

\date{Accepted, Received 2005 February 20; in original form }

\pagerange{\pageref{firstpage}--\pageref{lastpage}} \pubyear{2005}

\maketitle

\label{firstpage}

\begin{abstract}
A growing  number of recent observations have revealed that the Galactic
globular cluster (GC) $\omega$ Cen is not the only
GC that shows abundance spread in heavy elements (e.g., Fe).
In order to understand the origin of the Galactic GCs with heavy element abundance
spread (``HEAS''), we investigate the formation processes of massive GCs (MGCs) with 
masses larger than $10^{6} {\rm M}_{\odot}$ in gas-rich dwarf galaxies interacting and merging
with the very young Galaxy.
We find that massive and compact stellar clumps with masses larger than 
$10^{6} {\rm M}_{\odot}$, which can be regarded as progenitors of MGCs, 
can form from massive gas clumps  that are developed through merging of gaseous regions 
initially at different radii  and thus 
with different  metallicities.
Therefore it is inevitable that MGCs formed in dwarfs have HEAS.
The abundance spread in each individual MGC
depends on the radial metallicity gradient of the host  dwarf
such that it can be larger for the steeper metallicity gradient.
For example,  MGCs formed in a dwarf with a central metallicity of [Fe/H]$=-1.1$ and
the radial gradient of $\sim -0.2$ dex kpc$^{-1}$ can have the abundance spread
of $\Delta {\rm [Fe/H]} \sim 0.2$.
The simulated MGCs appear to be significantly flattened owing to their dissipative formation
from gas disks of their host dwarfs.
Based on these results, we discuss possibly diverse formation mechanisms for the Galactic
GCs such as M22, M54, NGC 2419, $\omega$ Cen, and Terzan 5.
\end{abstract}


\section{Introduction}

One of remarkable  recent developments in observational studies of the 
Galactic GCs is that most of the investigated GCs show varying degrees of chemical abundance 
spread: helium abundance spread in  $\omega$ Cen and NGC 2808
(e.g., Bedin et al. 2004; Piotto et al. 2007),  
a larger dispersion in
$s$-process elements abundances in  NGC 1851 (e.g., Yong \& Grundahl 2008; Milone et al. 2011),
abundance spread in light element  in ``normal'' GCs (e.g., Carretta et al. 2010a),
and HEAS in M22, Terzan 5, and NGC 2419 (e.g., Da Costa
et al. 2009; Ferraro et al. 2009;
Marino et al. 2009; Cohen et al. 2010). 
Lee et al. (2009) investigated  
color-magnitude diagrams
of stars
in $hk$-bands for the Galactic GCs and
suggested that
a significant fraction of the Galactic GCs have two different populations
with HEAS (but see also Marino et al. 2009; Carretta et al. 2010b  for
similar and different claims).

The origin of the HEAS in $\omega$ Cen
has been discussed by theoretical models in the context
of formation and evolution of stellar galactic nuclei in dwarfs 
(``the stripped nucleus scenario'';  e.g., Bekki \& Freeman 2003; Romano et al. 2010).
Since stellar populations of stellar galactic nuclei can have different
metallicities and ages owing to their possible formation
by merging of different stellar and gaseous
clumps (e.g., Bekki 2007),
it would  not be so  surprising that   
$\omega$ Cen 
is observed to  have HEAS 
owing to its possible origin from the stellar nucleus of its host dwarf.
Also, it would be possible that 
the formation process of other GCs with HEAS 
(e.g., Terzan 5) can be different from that of $\omega$ Cen 
(e.g., GC mergers;   Br\"uns \& Kroupa 2011; Bekki \& Yong 2011, BY11).  
Ideally speaking,  chemodynamical simulations  of GC formation based on a theoretical model
are essential
for discussing both the formation processes of GCs from gas
and the resultant internal chemical abundance spread within GCs 
in a fully self-consistent manner.
However no chemodynamical simulations on the origin of GCs with HEAS (other than $\omega$ Cen)
have been done so far.

The purpose of this Letter is to clearly show whether and how GCs with  HEAS 
can be formed by using chemodynamical simulations
of GC formation  processes based on a  GC formation scenario. 
We here consider that most of the Galactic GCs were formed in galactic building blocks
(e.g.,  Searle \& Zinn 1978), in particular, when they were interacting and merging
with the very young Galaxy ($>10$ Gyr ago).
We focus exclusively  on [Fe/H] spread ($\Delta {\rm [Fe/H]}$) in MGCs
with masses as large as or larger than $10^6 {\rm M}_{\odot}$ formed  within
their host dwarfs (i.e., galactic building blocks).

\begin{figure*}
\psfig{file=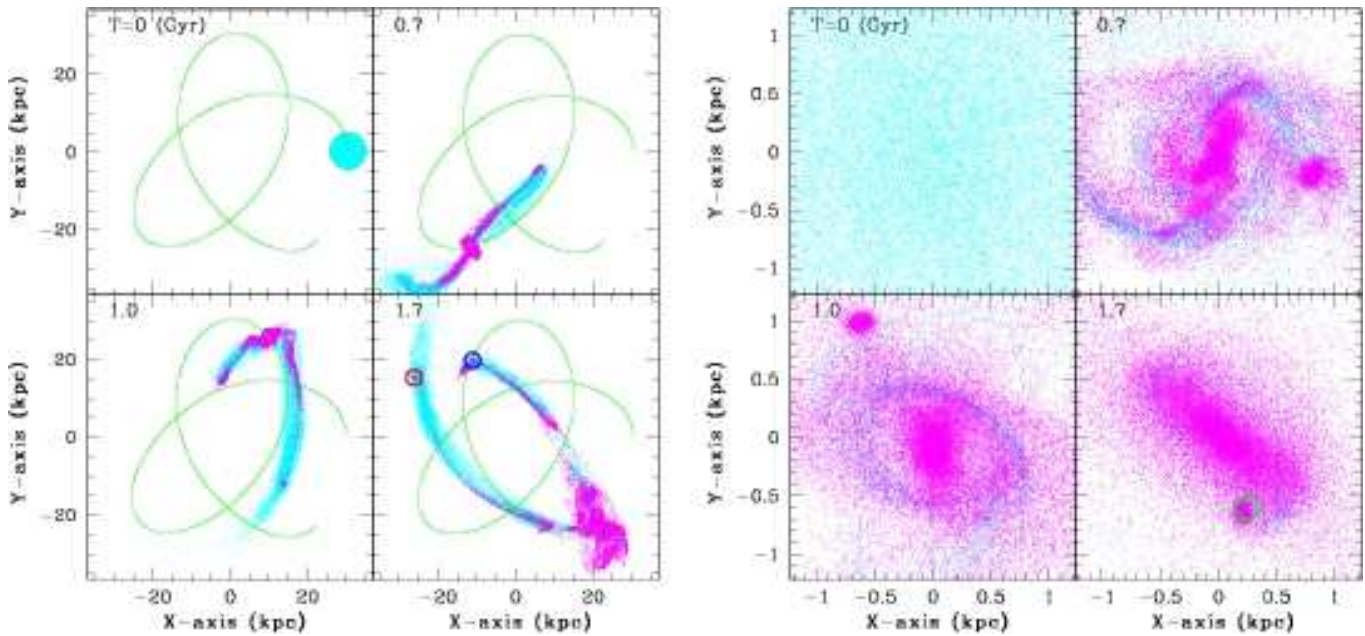,width=18.0cm}
\caption{
The time evolution of the mass distribution of the dwarf with respect to the Galactic center
(left four) and to the dwarf's center (right four) projected onto the $x$-$y$ plane for the 
standard model (M1). Gas and new stars are shown by cyan and magenta, respectively, 
and $T$, which is the time that has elapsed since the simulation started,
is shown in the upper left corner in units of Gyr for each panel. The green lines in the left
four panels represent the orbit of the dwarf. The locations of MGC1, 2, and 3 are indicated
by blue, red, and green circles at $T=1.7$ Gyr, respectively. 
}
\label{Figure. 1}
\end{figure*}

\section{The model}

We numerically investigate star formation and chemical evolution processes of a gas-rich dwarf
interacting and merging with the Galaxy in order to investigate whether MGCs composed of
many stellar particles  can be formed. 
In order to simulate chemodynamical evolution of the gas-rich dwarf in the young Galaxy,
we use both the latest version of GRAPE
(GRavity PipE,  GRAPE-DR), which is the special-purpose
computer for gravitational dynamics (Sugimoto et al. 1990),
and high-end PCs with GPU cards (GTX 580) 
and CUDA G5/G6 software package being
installed for calculations of gravitational dynamics.
We adopt our original GRAPE-SPH code (Bekki 2009)
which combines
the method of smoothed particle
hydrodynamics (SPH) with GRAPE for calculations of three-dimensional
self-gravitating fluids in astrophysics.

The dwarf galaxy is assumed to be strongly influenced 
by the fixed gravitational potential of the Galaxy in the preset study.
We adopt a  model of the Galaxy similar to  that used in BY11 in which
the Galaxy is assumed have
three components: a dark matter halo, a disk,
and a bulge. 
We consider that (i) dwarfs forming and hosting GCs interact or merge with the Galaxy in the very
early history of the Galaxy formation ($>10$ Gyr ago) and thus (ii) the Galaxy
has not yet fully developed its halo, disk, and bulge components at the epoch 
of dwarf merging.

We assume the following  logarithmic dark matter halo potential
for the Galaxy,
\begin{equation}
{\Phi}_{\rm halo}=v_{\rm halo}^2 \ln (r^2+d^2),
\end{equation}
where
$d$ = 12 kpc, $v_{\rm halo}$ = 131.5 km ${\rm s}^{-1}$ and
$r$ is the distance from the center of the Galaxy.
The gravitational potential of the Galactic disk is represented by
a Miyamoto-Nagai (1975) potential;
\begin{equation}
{\Phi}_{\rm disk}=-\frac{GM_{\rm disk}}{\sqrt{R^2 +{(a+\sqrt{z^2+b^2})}^2}},
\end{equation}
where $M_{\rm disk}$ = 1.0 $\times$ $10^{10}$ $M_{\odot}$,
and $a$ = 6.5 kpc, $b$ = 0.26 kpc,
and $R=\sqrt{x^2+y^2}$.
The disk mass  is much smaller than the one adopted in BY11, 
because we consider that only the first generation of the thin disk
is being built when the dwarf starts interacting/merging with the Galaxy ($>10$ Gyr ago).
We do not include the bulge in the young Galaxy, because we consider that the bulge 
can be formed much later from bar instability of disk stars.

The center of the Galaxy is always
set to be ($x$,$y$,$z$) = (0,0,0) whereas the initial location and velocity
of the dwarf are free parameters that can control the orbital evolution
of the dwarf.
The initial distance of the dwarf from the Galactic center
and the velocity are represented by $R_{\rm i}$ and $f_{\rm v}v_{\rm c}$,
respectively,
where $v_{\rm c}$ is the circular velocity at $R_{\rm i}$.
The inclination angle between
the  initial orbital plane and the $x$-$z$ plane (=Galactic disk plane)
is denoted as $\theta$. 
The initial spin of  the dwarf disk is specified by two angles,
$\theta_{\rm d}$ and $\phi_{\rm d}$, where
$\theta$ is the angle between the $z$-axis and the vector of
the angular momentum of a disk and
$\phi$ is the azimuthal angle measured from $x$-axis to
the projection of the angular momentum vector
of a disk onto the $x-y$ plane.
In the present study, we show the results of the models
with  $R_{\rm i}=35$ kpc,
$f_{\rm v}=0.5$,
$\theta=30^{\circ}$,
$\theta_{\rm d}=15^{\circ} $,
and $\phi_{\rm d}=45^{\circ}$.

\begin{table}
\centering
\begin{minipage}{85mm}
\caption{Description of the  model parameters for
the representative models.}
\begin{tabular}{cccccc}
{Model no.
\footnote{Only the model H1 is the higher resolution model with $N=10^6$.}}
& {$M_{\rm s, dw}$
\footnote{The total mass of a stellar disk in a 
dwarf   in units of $10^8 {\rm M}_{\odot}$.}}
& {$R_{\rm s, dw}$
\footnote{The initial size  of a stellar disk in a dwarf  in units of kpc.}}
& {$f_{\rm g}$
\footnote{The initial gas mass fraction ($M_{\rm g, dw}/M_{\rm s,dw}$).}}
& {$\alpha_{\rm d}$
\footnote{The radial metallicity gradient of the gas disk in a dwarf (dex kpc$^{-1}$).}}
& {$y_{\rm met}$
\footnote{The value of a chemical yield ($y_{\rm met}$)  is  not shown for models without 
chemical evolution. }} \\
M1 & 1.0  & 1.8 & 0.5 & $-0.2$  & - \\
M2 & 1.0  & 1.8 & 0.5 & $-0.1$  & - \\
M3 & 1.0  & 1.8 & 0.5 & $-0.05$  & - \\
M4 & 1.0  & 1.8 & 0.1 & $-0.2$  & - \\
M5 & 1.0  & 1.8 & 0.2 & $-0.2$  & - \\
M6 & 1.0  & 2.8 & 0.5 & $-0.2$  & - \\
M7 & 1.0  & 1.8 & 0.5 & $-0.2$  & 0.0005 \\
M8 & 1.0  & 1.8 & 0.5 & $-0.2$  & 0.002 \\
H1 & 1.0  & 1.8 & 0.5 & $-0.2$  & - \\
\end{tabular}
\end{minipage}
\end{table}

We adopt the same dwarf galaxy model as that in BY11: the dwarf has a ``cored'' dark matter
halo (Salucci \& Burkert 2000) and an exponential stellar disk. 
We describe the result of the models with the total stellar disk masses
($M_{\rm s, dw}$) being $10^8 {\rm M}$ in the present study: other models with different
$M_{\rm s, dw}$ will be discussed in our forthcoming papers. The mass-ratio of dark matter halo
to stellar disk in the dwarf is set to be 9 and thus the core radius ($a_{\rm dm}$) 
of the dark matter halo
is 1.5 kpc.
The stellar disk has
the total mass of $M_{\rm s,dw}$ and the size of $R_{\rm s, dw}$.
and the radial ($R$) and vertical ($Z$) density profiles of the stellar disk are
assumed to be proportional to $\exp (-R/R_{0}) $ with scale
length $R_{0} = 0.2R_{\rm s, dw}$ and to ${\rm sech}^2 (Z/Z_{0})$ with vertical scale height 
$Z_{0} = 0.04R_{\rm s, dw}$ , respectively.
In addition to the
rotational velocity caused by the gravitational field of disk
and dark halo components, the initial radial and azimuthal
velocity dispersions are assigned to the disc component according to
the epicyclic theory with Toomre's parameter $Q$ = 1.5.  The
vertical velocity dispersion at a given radius is set to be 0.5
times as large as the radial velocity dispersion at that point.

In the present study,
the dwarf also has an extended gas disk 
with the total mass of $M_{\rm g,dw}$ and the size of $R_{\rm g, dw}$ ($=2.5 R_{\rm s,dw}$).
The disk is assumed to  have an exponential profile with scale length  $0.4R_{\rm g, dw}$
and vertical scale height $0.04R_{\rm g, dw}$. The mass-ratio of gas disk to stellar one
($f_{\rm g}$) is an important free parameter that can control the formation processes
of MGCs in the present study. 
We consider that the dwarf disk initially has  a plenty of  cold gas
and accordingly  an isothermal equation of state
is adopted for the gas with a temperature ($T_{\rm g}$) of $160$ K.
Star formation
is modeled by converting  the collisional
gas particles
into  collisionless new stellar particles according to the algorithm
of star formation  described below.
We adopt the Schmidt law
with exponent $\gamma$ = 1.5 (1.0  $ < $  $\gamma$
$ < $ 2.0, Kennicutt 1998) as the controlling
parameter of the rate of star formation.

We allocate metallicity to each gas particle
according to its initial position:
at $r$ = $R$,
where $r$ ($R$) is the projected distance (in units of kpc)
from the center of the disk, and thus the metallicity of the gas is given as:
\begin{equation}
{\rm [Fe/H]}_{\rm r=R} = {\rm [Fe/H]}_{\rm d, r=0} + {\alpha}_{\rm d} \times {\rm R}. \;
\end{equation}
We consider  that (i) the slope ${\alpha}_{\rm d}$ is a free parameter and
(ii) ${\rm [Fe/H]}_{\rm d, r=0}$ is determined by the total stellar mass of the dwarf.
Accordingly we adopt [Fe/H]$=-1.1$ as a reasonable value for the adopted $M_{\rm s, dw}$,
and the observed  value  of ${\alpha}_{\rm d} \sim -0.2$
for dwarf spirals (Hidalgo-G\'amez et al. 2010) should be a reference value for the radial metallicity
gradient. The chemical yield $y_{\rm met}$ and the return parameter
$R_{\rm met}$ are key parameters that control chemical evolution of dwarfs.
We investigate models with (i.e., $y_{\rm met}>0$ and a fixed $R_{\rm met}=0.3$) 
and without ($y_{\rm met}=0$)
chemical enrichment by  supernovae.

In order to find massive stellar clumps with masses ($m_{\rm ns}$)  
as large as or larger than  $10^6 {\rm M}_{\odot}$,
we investigate the total number of new stellar particles  ($N_{\rm nei}$) within 200pc 
around each new stellar particle.  If $N_{\rm nei}$ around a new stellar particle 
is larger than a threshold value $n_{\rm thres}$, then
we regard the particle as being  within a massive and compact stellar clump.
Since the particle is not necessarily at the center of the clump, we newly determine the 
center of the clump using the position data of the neighboring particles around
the particle.  For the newly determined center of the clump,
we estimate the total mass of the clump within the central 200pc.
We consider that $n_{\rm thres}=1000$ is reasonable, firstly because 
only massive clumps ($m_{\rm ns} \ge 5 \times 10^5 {\rm M}_{\odot}$ 
for $f_{\rm g}=0.5$) can be selected
and secondly because the selected clumps also show high stellar densities.

The selected massive clumps are significantly larger and more massive than the present GCs
and thus should be regarded as progenitors of GCs (i.e., not literally  genuine GCs).
Recent numerical studies of multiple stellar populations in GCs have suggested that  the original GC 
{\it should be}  much larger and
more massive than the present ones (e.g., D'Ercole et al. 2008; Bekki 2011).
Therefore, the adopted assumption that massive stellar clumps can be progenitors of GCs 
is reasonable and realistic.  The final massive stellar clumps are referred to
as MGCs for convenience in the present study.
Owing to the introduction of a  gravitational softening length for new stars,
the present simulations are not suitable for investigating the long-term internal 
dynamical evolution of GCs. We will discuss the long-term dynamical evolution
processes of GCs after their formation in gas-rich dwarfs in our forthcoming paper
by using N-body codes suitable for GC internal dynamics.
 Table 1 summarizes nine representative models for which the results are discussed in the present
paper. We mainly describe the results of the ``standard'' model (M1)  in which
$f_{\rm g}=0.5$, $\alpha_{\rm d}=-0.2$, and
chemical enrichment is not included (i.e., $y_{\rm met}$=0).
This is mainly because we need to more clearly demonstrate
the origin of HEAS in GCs.
The total particle numbers 
for the dark matter ($N_{\rm dm}$),  the stellar disk  ($N_{\rm s}$),
and the gas one ($N_{\rm g}$)
used in most of the present models 
are 200,000,  200,000, 100,000, respectively. 
The gravitational softening lengths ($\epsilon$)  for the dark matter particles,
the stellar ones, and the gaseous ones are 
128pc, 20pc, and 20pc, respectively.
In order to confirm that the present results do not depend on the adopted numerical
resolution, we also run a high-resolution model (H1) in which 
$N_{\rm dm}$=400,000
$N_{\rm s}$=400,000
and $N_{\rm g}$=200,000.

\begin{figure}
\psfig{file=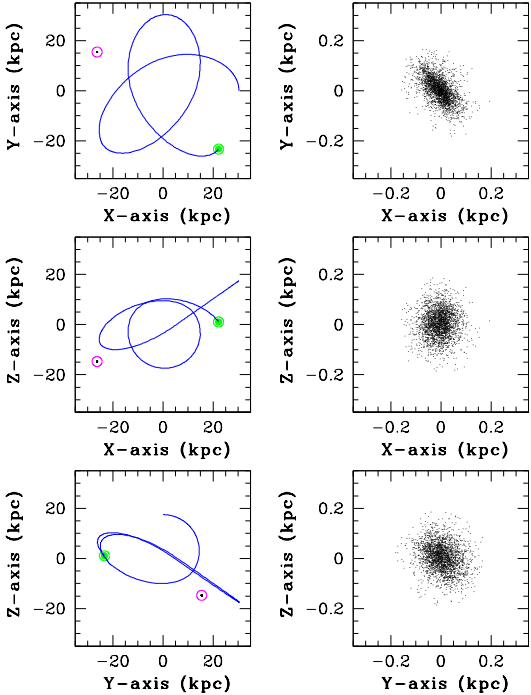,width=7.0cm}
\caption{
The orbits of the dwarf (left three) and the mass distributions of MGC2 (right three) 
projected onto the $x$-$y$ (top), $x$-$z$ (middle), and $y$-$z$ planes (bottom)
in the standard model (M1). For comparison,
the locations  of the dwarf and MGC2 at $T=1.7$ Gyr are shown by green and magenta circles,
respectively. Only new stars formed from gas are plotted in the right three panels.
}
\label{Figure. 2}
\end{figure}

\begin{figure}
\psfig{file=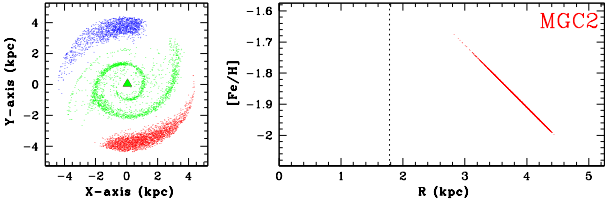,width=8.0cm}
\caption{
The original location of new stars of MGC1 (blue),
2 (red), and 3 (green)  within the initial dwarf
disk ($T=0$ Gyr)  projected onto the $x$-$y$ plane
(left) and the initial locations 
of original gas particles for  new stars of MGC2 on the $R$-[Fe/H] plane
(right).   The green triangle  represents the final location of MGC3 
within the host at $T=3.4$ Gyr (i.e., evolution into nucleus). 
The vertical dotted line in the right panel represents the size of the stellar disk
of the host dwarf ($R_{\rm s, dw}$). 
}
\label{Figure. 3}
\end{figure}

\begin{figure}
\psfig{file=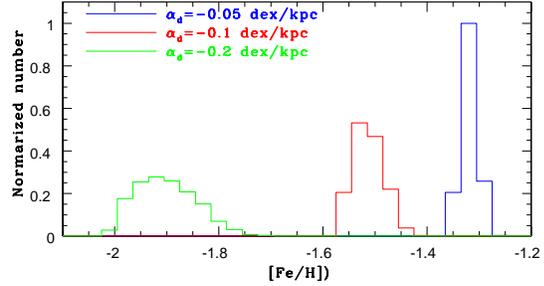,width=7.0cm}
\caption{
The [Fe/H] distributions of new stars in MGC2 for 
the models M3 with $\alpha_{\rm d}=-0.05$ (blue),
M2 with $\alpha_{\rm d}=-0.1$ (red),
and M1 (i.e., the standard model) with $\alpha_{\rm d}=-0.2$ (green).
}
\label{Figure. 4}
\end{figure}

\section{Results}

Fig. 1 shows the orbital evolution of the dwarf and the time evolution of the mass distributions
of gas and new stars in the standard model (M1).  
During the first pericenter passage ($R_{\rm p} \sim 14$ kpc) at $T \sim 0.3$ Gyr,
the gas disk is strongly compressed by the  tidal field of the Galaxy to form high-density gaseous
regions. As a result of this, a number of massive clumps composed mostly of gas and new stars 
can be formed within the central $\sim 1$ kpc of the dwarf and the star formation rate
becomes higher ($\sim 0.08 {\rm M}_{\odot}$ yr$^{-1}$).
After the pericenter passage, a significant fraction
of gas and new stars are tidally stripped and the stripped matter can form leading
and trailing streams along the orbit of the dwarf ($T=0.7$ Gyr).  Some of the more massive clumps
can sink into the central region of the dwarf owing to dynamical friction so that
they can become stellar  galactic nucleus. At the final time step ($T=1.7$ Gyr),
only one massive and compact clump (identified as MGC3)
with $m_{\rm ns}=1.2 \times 10^6 {\rm M}_{\odot}$ 
(originally $1.6 \times 10^6 {\rm M}_{\odot}$ at $T=1.0$ Gyr) can be clearly
seen within the central 1 kpc of the dwarf.  
 We confirm that this MGC3 
can sink into the dwarf's center within the next $\sim 1$ Gyr 
to become nucleus: it can be identified as an isolated GC
only after the complete destruction of its host 
by the (older) Galaxy with the much larger mass.

Two stellar clumps 
can be developed from gaseous tidal tails (or outer spiral arms) during
the tidal interaction of the dwarf with the Galaxy. Fig. 1 shows that the two clumps
(i.e., MGC progenitors) 
can be stripped from their host dwarf and consequently  located  
within the gaseous tails ($T=1.7$ Gyr).
These two clumps are identified as  
MGC1 with  $m_{\rm ns}=7.5 \times 10^5 {\rm M}_{\odot}$
and MGC2 with  $m_{\rm ns}=1.3 \times 10^6 {\rm M}_{\odot}$.
They can be finally regarded
as ``isolated'' GCs 
owing to their large distances ($>40$ kpc) from the dwarf.
Fig. 2 shows  that
owing to its formation
from the outer gas disk through gaseous dissipation,
MGC2 has a  significantly flattened shape in the $x$-$y$ projection.
This flattened shape can be seen in most of the simulated MGCs in the present study.

The formation of MGCs can be severely suppressed in models (M4 and 5)
with lower $f_{\rm g}$ ($ \le 0.2$), which implies that MGC formation
is possible only when their host dwarfs have a plenty of cold gas.
Therefore the formation processes of MGCs in gas-rich disks and tidal tails 
described above can be essentially the same as 
those of massive stellar and gaseous clumps discussed in previous papers 
(e.g., Noguchi 1999; e.g., Bournaud et al. 2007) in which
$f_{\rm g}$ is a fundamentally  important parameter 
for dynamical evolution of gas-rich galactic disks.
MGCs can be formed even in the low surface brightness model M6 with larger
$R_{\rm s, dw}$ (=2.8 kpc), which implies that mass densities of stellar disks in dwarfs
are not so important as $f_{\rm g}$ for MGC formation.

Fig. 3 clearly shows that new stars of MGC2 originate from gaseous regions
at different locations within the dwarf and thus with different [Fe/H] owing
to the adopted metallicity gradient of the gas disk. This result demonstrates that
some Galactic GCs have HEAS because they formed from massive gaseous clumps developed
from merging of different gaseous regions with different initial [Fe/H].  
However,  Fig. 4 shows that the degree of HEAS ($\Delta$[Fe/H]) depends strongly on
$\alpha_{\rm d}$ such that $\Delta$[Fe/H] is larger ($\sim 0.2$ dex) for the steeper
metallicity gradient (i.e., the larger absolute magnitude of $\alpha_{\rm d}$).

 Fig. 5 shows that 
a younger population can have 
[Fe/H]$\sim -1$ in MGC2 
in the model with chemical evolution (M7),
though the mass fraction the population with the age less than 0.2 Gyr
is only $\sim 0.05$. Although the prolonged star formation
can be seen,  the vast majority of stars are formed in a burst at
the epoch of massive clump formation.
Furthermore, there appears to be an age-metallicity relation that
younger stars are more metal-rich in MGC2. 
It should be stressed that the relation
can depend on how the supernova ejecta can influence
the later star formation within cluster-forming massive
gas clumps, which needs to be investigated by more sophisticated
models in our future works: 
The [Fe/H] spread could be overestimated owing to
the adopted instantaneous recycling approximation.
The mean metallicities and internal [Fe/H] spread 
of MGCs are  larger 
in models with larger $y_{\rm met}$ (e.g., M8) in the present study.
The physical properties of MGCs (e.g., masses and locations)
are not different 
between models with different numerical resolutions (M1 and H1).
 
\begin{figure}
\psfig{file=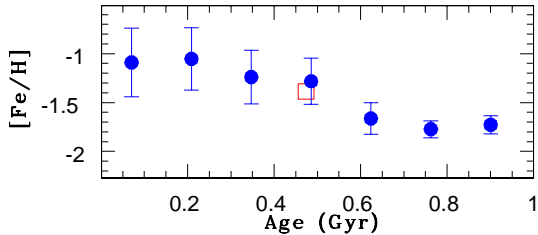,width=7.0cm}
\caption{
The age-metallicity relation for MGC2 in the model M7 with chemical evolution 
($y_{\rm met}=0.0005$). 
Here ``ages'' ($t_{\rm age}$) 
of stars are estimated at $T=1.7$ Gyr (i.e., the final time step of the simulation):
$t_{\rm age}=0$  means star formation at $T=1.7$ Gyr.
The mean value and dispersion of [Fe/H] 
at each age bin are shown by
a filled blue circle and an error bar, respectively.  The red square represents the mean
age and metallicity of MGC2.
}
\label{Figure. 5}
\end{figure}
\section{Discussion and conclusions}
The present chemodynamical study has first demonstrated that MGCs with HEAS
can be formed from massive gas clumps 
developed from merging of different gaseous regions with different metallicities. 
Therefore, MGCs with
HEAS can be formed even without chemical enrichment by supernovae during their formation.
The present study suggests that more massive GCs are more likely to have a larger degree of
HEAS owing to their formation from gas clouds developed from merging between a larger number
of different gaseous regions with different metallicities. The low-mass GCs can be formed
from single gas clouds so that they are likely to have no/little HEAS. The present study thus
suggests that the origin of HEAS in some Galactic GCs can be closely associated with
HEAS of ISM in their host dwarfs.

The large metallicity spread of various elements and possible age variation
in $\omega$ Cen 
(e.g., Sollima et al. 2005) can be consistent with
the stripped nucleus scenario.  However, this does not necessarily mean that all of the 
Galactic GCs with HEAS (e.g., NGC 2419, M22, and Terzan 5) 
were formed from stripped nuclei of dwarfs. 
A GC with distinct two peaks in the [Fe/H] distributions of the stars
might well be consistent with GC merging with different [Fe/H]
(e.g., Br\"uns \& Kroupa 2011; BY11).
A GC with a smaller  degree of HEAS yet no clear bimodal [Fe/H] distribution might well
form from massive gas clumps of the host dwarf 
and then be stripped from the dwarf without sinking into
the center (i.e., without  becoming the stellar nucleus) to finally become the Galactic halo  GC. 
We lastly suggest that  nucleation,  GC merging,  and merging of gas clouds with different
[Fe/H] can  be all promising mechanisms  for the 
formation of the Galactic GCs with different degrees of HEAD.  
\section{Acknowledgment}
I am   grateful to the anonymous referee for constructive and
useful comments that improved this paper.

\end{document}